\documentclass[acmjacm,hyperref]{acmtrans2m}
\usepackage{graphicx}
\usepackage{amsmath}
\usepackage{amsfonts}
\usepackage{amssymb}
\usepackage{enumerate}

\title{User interfaces and data entry with real time inverse arithmetic coding}
\author{PATRICK KAIFOSH \\ Current email: pwk2108@columbia.edu }

\begin{abstract}
This paper introduces real time inverse arithmetic coding and user interfaces based thereupon.
The main idea is that information-efficient data entry can be achieved by ensuring that each input's associated display space and ease of selection are at all times related to the input's probability of being selected.
As with data entry based on inverse arithmetic coding, the layout initially depends on the probabilities of the possible inputs;
however, real time inverse arithmetic coding differs in that the user's actions are interpreted not to navigate this probability distribution
but rather to modify it according to a learned update rule, which approximates the conditioning of the probability distribution on the user's actions.
Potential applications of real time inverse arithmetic coding include text entry, file browsing, integrated multi-program user interfaces, assistive technologies for users with movement disabilities, and ergonomic input methods.
\end{abstract}

\category{H.5.2}{Information Interfaces and Presentation}{User Interfaces}[Interaction styles \and Theory and methods \and User-centered design]
\category{K.4.2}{Computer and Society}{Social Issues}[Assistive technologies for persons with disabilities]
\category{K.8.1}{Personal Computing}{Application Packages}[Word processing]

\terms{Design, Human factors, Theory}

\keywords{Data entry, inverse arithmetic coding, user interface}

\begin{document}

\setcounter{page}{1}

\maketitle

{\makeatletter
\gdef\@runningfoot{}
\gdef\@firstfoot{}
}

\section{Introduction}\label{Introduction}
\begin{figure}
\centering
\includegraphics[width=0.9\textwidth]{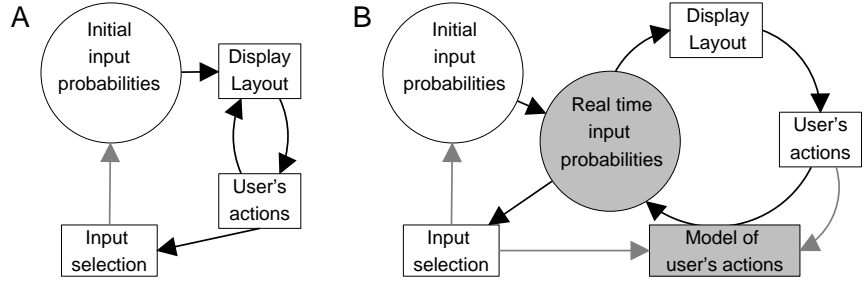}
\caption{\label{schemas} (A) IAC and (B) RTIAC data entry schemes. 
Black arrows indicate immediate effects; gray arrows indicate slower learning effects.
In both schemes, the initial input probability distribution determines the initial layout and thus influences the user's actions,
and the statistics of the user's selections are used to refine the initial input distribution.
In IAC, the user selects the input by navigating amongst the displayed initial input distribution.
In RTIAC, the user's actions modify the input distribution in real time according to a learned update rule, and the display is modified accordingly;
an input is selected once its probability exceeds a given threshold.}
\end{figure}

Concepts from information theory are generally applicable to all forms of communication, including that between computers and their users.
For example, arithmetic coding has been particularly fruitful in inspiring \textit{Dasher}, an information-efficient data entry device based on inverse arithmetic coding (IAC) \cite{Ward2000}.
The key concept of IAC data entry is that,
if each input's selection ease (i.e.\ the paucity of information required to make the selection) and proportion of display space are related to its probability of selection,
the rate of information transfer between the computer and human is enhanced in both directions: 
the computer can more rapidly convey information about possible inputs to the user, and the user can more rapidly convey information about choices from among these inputs.
Applicable not only to text entry, IAC is a general concept (Fig. \ref{schemas}A) that can be used in designing and improving interfaces through which users communicate data and instructions to machines.

However, IAC methods are currently limited in that the probability distribution and consequent layout are never updated to reflect the user's actions, which instead serve only to navigate within the original layout of possible inputs.
This limitation prevents IAC from fully optimizing the extent to which the probability of an input is reflected in its proportion of display space and selection ease, and the rate of information transfer between the user and the computer is reduced accordingly.
The situation can be improved if the probability distribution used for IAC is updated in real time in response to the user's actions;
this concept of real time inverse arithmetic coding (RTIAC), depicted schematically in Fig.\ \ref{schemas}B, and its possible implementations and applications are the focus of this paper.

\section{Inverse arithmetic coding data entry}
\subsection{Overview}
Data entry using IAC, introduced by \citeN{Ward2000}, is based on arithmetic coding of strings \cite{Witten1987}.
Given an ordered alphabet $\mathcal A$, a probability distribution on $\mathcal A^*$ (the set of finite strings over $\mathcal A)$, and a termination symbol $\tau$,
each string $s\in A^*\cup(A^*\cdot \tau)$ is assigned an interval $I_s=[a_s,b_s)$ according to the following rules:\footnote{
Notation: $\bigsqcup$ denotes a disjoint union, $s\cdot s'$ the concatenation of $s$ with $s'$, $P(s)$ the probability of string $s$, and $A^*\cdot \tau$ the set $\{s \;|\; s=s'\tau$ for some $s'\in \mathcal A^*\}$.}
\begin{enumerate}[(a)]
 \item $\bigsqcup_{a \in \mathcal A\cup\{\tau\}} I_a = [0,1)$,
 \item for each $s\in \mathcal A^*$, $\bigsqcup_{a \in \mathcal A \cup \{\tau\}} I_{s\cdot a} = I_s$,
 \item for each $s\in\mathcal A^*$ and $a,b\in\mathcal A\cup\{\tau\}$, if $a < b$, $x\in I_{s\cdot a}$, and $x'\in I_{s\cdot b}$, then $x < x'$,
 \item $b_{s} - a_{s} = \sum_{s'\in A^*} P(s\cdot s')$.
\end{enumerate}
Essentially, the unit interval $[0,1)$ is divided into disjoint subintervals each corresponding to the possible first characters,
each subinterval is then recursively divided into further disjoint subintervals corresponding to the subsequent character,
the intervals are ordered according to the dictionary ordering of the corresponding strings,
and the length of the interval for each string in $\mathcal A^*$ is equal to the combined probability of all strings beginning with the corresponding string.
These rules define a unique mapping from strings to interval subsets of $[0,1)$.

In IAC data entry, as realized in the text entry program \textit{Dasher} (Fig.\ \ref{dasher}),
the user navigates a visualization of the unit interval by cursor-controlled translation and zooming until the interval corresponding to the desired input is found \cite{Ward2000}.
The sizing of intervals in relation to the probabilities of an input string provides two advantages that increase the rate of data entry: 
First, likely entries are easy to find and can be selected quickly without the user having to convey much information.
Second, since the probability of selection is uniformly distributed on the unit interval and thus over the vertical axis,
the user is equally likely to move the cursor to each height on the screen and thus communicates information to the computer at a relatively rapid rate;
the computer can then discern the user's choice more quickly than if the user were most likely to select from a more restricted proportion of the screen.
With these advantages, the \textit{Dasher} program allows for reasonably fast text entry with a low error rate when controlled by either mouse \cite{Ward2000} or eye tracker \cite{Ward2002}.

\begin{narrowfig}{5in}
\begin{tabbing}
\includegraphics[width=0.4\textwidth,clip=true,trim=0 1.1cm 0 2.05cm]{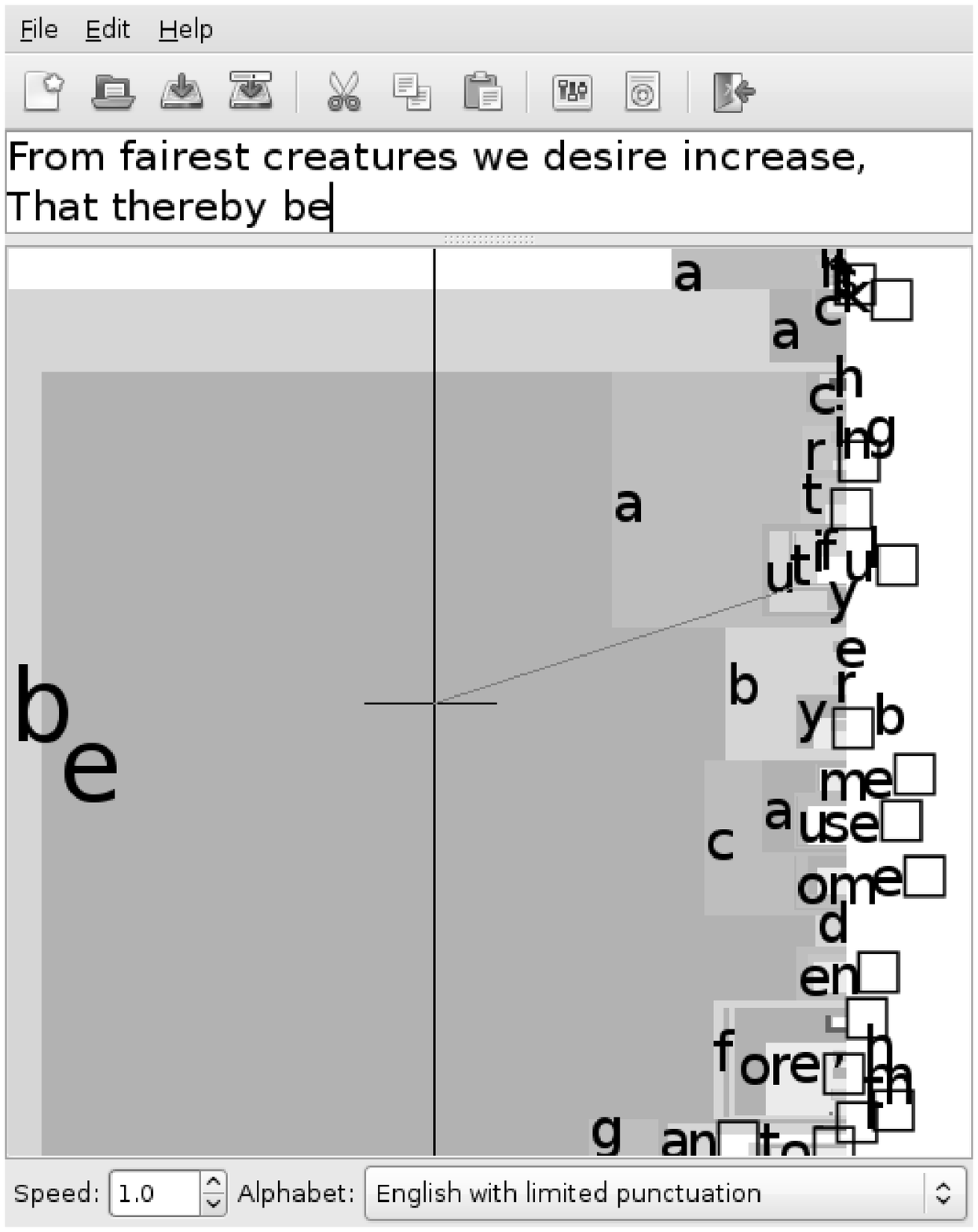}
\end{tabbing}
\caption{The \textit{Dasher} text entry program \cite{DasherProject}.
Each character is associated with a square, and strings are displayed by nesting the square of each subsequent character within that of the previous one.
Placement of the cursor above or below the center causes the display to scroll in the respective direction;
placement of the cursor to the right of center causes the display to zoom in, while placement to the left of center causes the display to zoom out.
Movement of the cursor directs the end of the line toward the desired character string.
Essentially, the user navigates by pointing with the cursor at the desired input.}
\label{dasher}
\end{narrowfig}

\subsection{Limitations}\label{limitations}
IAC data entry is limited by the fact that the user's actions serve only to locate the desired input within the visualization of the prior distribution;
such a fixed navigation scheme prevents the computer from attaining the full information provided by the user's actions and maintaining a display layout that accurately reflects the input probability distribution.
As a result, the optimal conditions for rapid information transfer and data entry deteriorate once the user begins navigating the IAC interface to select an input.

These limitations are best seen through examination of the \textit{Dasher} navigational scheme,
according to which the vertical component of the cursor position causes traversing of the probability distribution while the horizontal component causes zooming in and out (Fig.\ \ref{dasher}).
The combined effect of the horizontal and vertical motions is that the screen continuously zooms in on the region to which the cursor is pointing.
The proportion of display given to each input is uniformly scaled within a certain region near the cursor and vanishes outside of that region.
However, there is still a non-zero probability that the user will select an input that has fallen outside of the displayed region due to earlier erroneous cursor movements.
This particular discrepancy between the input probabilities and the display layout requires that the interface allocate specific cursor positions to zooming out.
Although the probability of the desired input being outside the displayed region is typically much less than half, all of the cursor positions to the left of the display's center correspond to zooming outwards.
Other discrepancies include that uniform zooming does not reflect the higher likelihood that the user will pick the input coincident with the cursor than the input either above or below it, and that the rate of zooming as a function of the cursor's horizontal component will not necessarily reflect the proximity of the cursor's vertical component to the desired input.

\section{Real time inverse arithmetic coding}
\subsection{Main concept}
The limitations of IAC data entry arise from the fact that,
while the arithmetic coding scheme may faithfully reflect the probability distribution of character strings prior to any user input, it less accurately reflects how this probability distribution is modified by the user's actions.
Thus, although IAC has certain optimality properties at the initial time, these advantages are diminished as time progresses.

For the user interface to maintain optimality throughout time,
each input's amount of display space and ease of selection must always reflect the current input probability distribution.
Thus, while IAC data entry uses only a single initial probability distribution,
an improved approach requires a time dependent probability distribution with its frequency function given by
\begin{align}
 p(x,t) = p(x|I_t),
\end{align}
where $p(x|I_t)$ denotes the conditional probability that the desired input is $x$ given the user's actions $I_t$ up to the current time $t$.
At each instant, this conditional distribution can be used to determine the current layout just as the initial probability distribution determines the initial layout for IAC data entry.
Thus, as diagrammatized in Fig.\ \ref{schemas}B, the user's actions do not navigate within the visualization of the probability distribution but rather serve to refine the probability distribution underlying this visualization.
Methods following this approach will henceforth be referred to as real time inverse arithmetic coding (RTIAC).

Switching the focus from the prior probability distribution used in basic IAC schemes to the time-dependent conditional distribution discussed above allows for development of user interfaces with better information theoretic properties.
Specifically, the full information contained in the user's action is used by the interface.
As a result, the proportion of display area and ease of selection can remain related to each input's probability throughout time.

\subsection{Interface designs}
The RTIAC concept is not specific to any particular user interface.
Nevertheless, since the usefulness of RTIAC depends on effective interface designs, a necessarily incomplete discussion of this topic is warranted.

\subsubsection{Display layouts}\label{sec:layouts}
\begin{acmtable}{\textwidth}
\centering
\begin{tabular}{c|cccc}
    layout & linear & circular & prop.\ area & tree \\
\hline
area scaling & $p(x,t)^2$ & $p(x,t)^2$ & $p(x,t)$ & $\sim p(x,t)$\\
branching/nesting    & left-to-right & radial & inwards & along branches \\
ordering    & top-to-bottom & left-to-right & rows/2D & varies\\
cursor angles used & $90^\circ$ & $360^\circ$ & $360^\circ$ & $360^\circ$ \\
dimensions indicating choice & 1 & 1 & 2 & 2
\end{tabular}
\caption{Comparison of the linear, circular, proportional area, and tree display layouts.}
\label{tab:layouts}
\end{acmtable}

For user interfaces involving visual displays and continuous user actions (e.g. cursor movements),
the ideal display layout satisfies the following properties:
each input's display region size and selection ease are proportional to its probability,
these regions change continuously in a way that the user can easily follow, 
the user is induced to make a large variety of actions with even probability to maximize the rate at which information is communicated to the computer, and
the natural organization of the data being entered is reflected visually.
The layout schemes presented below each have different strengths and weaknesses in satisfying these requirements (Table \ref{tab:layouts});
none is optimal in all regards.

\begin{figure}
\centering
\includegraphics[width=\textwidth]{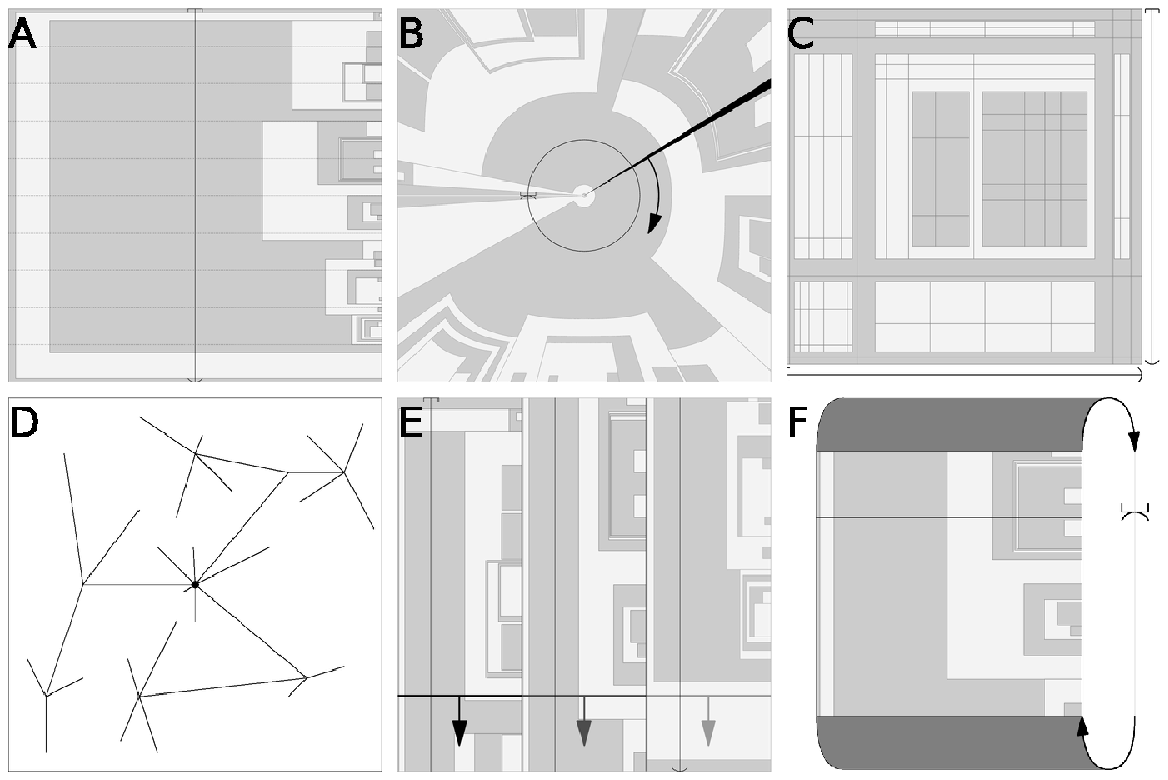}
\caption{Display layouts. 
A. Linear layout with dotted lines indicate regions associated with each of ten discrete inputs. 
B. Circular layout with single indicator for data entry with a single timed action. 
C. Proportional area layout. 
D. Tree layout.
E. Folded linear layout for data entry with three discrete timed actions.
F. Linear layout scrolling across a smaller display with a fixed indicator for timing-based input.
The arrows indicate direction of motion, and the symbols ` [ ' and ` ) ' indicate the beginning and end of each representation of $[0,1)$.}
\label{layouts}
\end{figure}

The linear layout (Fig.\ \ref{layouts}A) is that developed for \textit{Dasher} \cite{Ward2000}.
The unit interval is stretched along the vertical axis of the display,
and inputs are given rectangles with heights and widths both proportional to their probability.
The linear layout is amenable to data that are naturally ordered and tree-structured; 
such data include character strings, file directories, and locations within documents with parts, chapters, sections, and subsections.
Entries are ordered linearly along the vertical axis, and the tree branches along the horizontal axis.
For text entry, the left-to-right (or alternatively top-to-bottom) branching of the tree is advantageous for those used to reading character words in this orientation.
A major drawback of this scheme is that the displayed inputs can become quite crowded for two reasons:
the possible inputs are displayed only along the right quarter of the screen perimeter,
and the area allotted to each input scales with the square of its probability rather than the probability itself.
Information flow from the user is restricted by the fact that the cursor will be positioned predominantly within 45 degrees of the right horizontal semi-axis.

The circular layout (Fig.\ \ref{layouts}B) corrects this latter problem by casting the unit interval into a circle\footnote{
     To fit a standard rectangular display, the circle must be distorted. For simplicity, the discussion in this paragraph is presented in terms of the circle prior to distortion.} 
about the display's center so that the user is equally likely to move the cursor toward each angle.
To each input string $x$ is assigned the region outside an arc traversing an angle of $2\pi p(x,t)$ and with radius dependent on $p(x,t)$.\footnote{
     In Fig.\ \ref{layouts}B, the radius $r(p(x,t))$ is given by the inverted parabola $r(p) = c_0 + \sqrt{c_1+c_2p}$, with the coefficients $c_i$ chosen such that $r(0)=0$, $r(1)=1$, and $r'(0)=2\pi$. The last condition on the derivative sets the rectangular region's aspect ratios to unity in the limit of small probabilities.
     The widths of the rectangular regions in Fig.\ \ref{layouts}E-F have been calculated similarly.}
Although the area of each input's region still scales (in the limit of small probabilities) with the square of the input's probability, the scaling coefficient is improved on account of the larger screen perimeter.
The circular layout maintains all the advantages of the linear layout, except for the consistent left-to-right (or top-to-bottom) directionality; 
nevertheless, a consistent radial directionality is maintained.

In the proportional area layout (Fig.\ \ref{layouts}C), each input has a region whose area is proportional to the input's probability.
It is simplest to maintain this condition throughout time if one assumes that the user indicates the $x$ and $y$ coordinates of the desired input independently of each other;
this assumption, which can be generalized for three-dimensional displays, ensures that regions that are initially rectangular remain so.
Such rectangular shapes are convenient for displaying text and images in a manner that is easily viewed.
In improving the relationship between probability and region area, the proportional area layout loses its suitability for displaying a simple linear ordering, which must be folded into either rows or columns or lost completely.
However, in cases where the user must select pairs of two statistically independent input characteristics (e.g. points in two dimensions, independent and dependent variable values from an experiment, musical note pitches and durations), the proportional area layout can suit the natural ordering by having one axis correspond to the first element of the pair, and the other axis select the second; 
this is the situation in Fig.\ \ref{layouts}C.
Tree organization of the inputs is reflected in the nesting of some regions within others.

When the possible inputs are organized hierarchically but otherwise unordered, they can be visualized as a tree rooted at the center of the display (Fig.\ \ref{layouts}D), with nodes being displayed only when their associated inputs have probabilities exceeding some threshold.
Various algorithms can be used to position the nodes such that the each input's probability is approximately proportional to the display area nearest to the associated node and its descendents.
One possibility have simulate a dynamical system in which nodes repel each other to an extent dependent on their associated inputs' probabilities and in which edges provide an attractive force between between the nodes they connect.

In the proportional area and tree layouts, both directions of cursor movement tend to indicate the most probable input selection, since either each region or each node respectively has a position in two dimensions.
This situation contrasts with the linear and circular layouts in which the most likely input tends to be indicated primarily by the vertical and circumferential motions respectively, while the second dimension tends to provide information about the variance of this input selection.

\subsubsection{Discrete actions}\label{sec:single_key}
RTIAC can also be implemented with user interfaces in which the user makes discrete rather than continuous actions.
To allow a user to communicate with the timed performance of just a single action, such as the pressing of a button,
an indicator continually traverses the visualized unit interval can be added to the linear or circular layout (Fig.\ \ref{layouts}B).
The user attempts to signal when the indicator coincides with the desired input, and after each signal the input probability distribution is updated based on the statistics of the user's timing errors;
as well, the indicator's scrolling speed could be adjusted until the rate of information transfer were maximized.
Since only one dimension is used for indicating the selection choice, it may be advantageous to fold the linear layout into columns (Fig.\ \ref{layouts}E) in order to extend the axis along which selection is made.
Alternatively, with the rectangular layout, both dimensions can be used in selection provided that the axis scanned by the indicator alternate with each signaling event.
For small displays, it may be advantageous to scroll the input choices rather than move the indicator so that the whole visualization of the input probability distribution need not be displayed simultaneously (Fig.\ \ref{layouts}F).

When multiple discrete actions are possible, the preferable interface design will generally depend on the number of distinct actions.
For example, if just a few actions are available, having one moving indicator correspond to each action (Fig.\ \ref{layouts}E) may be desirable.
Whereas if many distinct actions are possible, permanently associating each action with a fixed subset of the unit interval (Fig.\ \ref{layouts}A) may be more effective.
With this latter method, which may also be preferred by users for whom precise timing is difficult, correct application of RTIAC implies that each action is interpreted not as selecting the associated subset of $[0,1)$ but rather, since the user makes mistakes, as increasing the probability that the input is in that subset.
This distinction is especially important when a great variety of discrete actions are possible
because mistakes should occur more frequently under such circumstances.

\subsection{Learning considerations}
Without loss of generality, the learning task can be discussed in the case that the user is specifying a number $x$ chosen uniformly at random from the unit interval.
At each time $t$, one can consider the transformed variable $y(x,t) = \int_0^x \hat p(x,t)$, where $\hat p(x,t)$ denotes the estimated probability of input $x$ at time $t$.
The learning task is then to determine $p(y,t+\delta t) = p(y|I_{t+\delta t})$, the probability density that the transformed variable was $y$ at time $t$, conditional on the user's actions up to time $t+\delta t$.
Since the user acts in response to the displayed position of the desired input, which is solely dependent on the transformed variable,
the transformation from $x$ to $y(x,t)$ makes the density $p(y|I_{t+\delta t})$ invariant under temporal translations and dependent only on the user's more recent actions.
If the probability distribution is conditioned on $I_{t+\delta t}^\tau$, the user's actions within only the most recent $\tau$ time units, instead of on $I_{t+\delta t}$,
then the dimensionality of the learning problem is reduced and the learning algorithm can be presented with identically distributed examples at a rapid rate.\footnote{
     Although examples can be generated arbitrarily often, there is limited value in generating many examples in a time interval over which the user's actions are strongly coherent.}
The choice of the time interval length $\tau$ is a trade-off between retaining more information and reducing the dimensionality of the probability density function that must be learned.

The most natural learning approach begins with recording the data pairs $(y,I_{t+\delta t)}^\tau$ of the values ultimately selected and the user's actions.
Kernel density estimation can then be applied to obtain a smooth joint density $p_s$,
which can be reweighted as $\hat p(y,I_{t+\delta t}^\tau) = p_s(y,I_{t+\delta t}^\tau) / p_s(y)$
to ensure that the marginal density $\hat p(y)$ is independent of $y$ in accordance with the definition of this transformed variable.
The RTIAC procedure can then use the estimated conditional probability density $\hat p(y|I_{t+\delta t}^\tau)$.

Although non-parametric methods, such as that above, reduce the bias of the estimated conditional probability density,
restriction of the hypothesis space provides allows for both faster learning (i.e.\ reduced variance) \cite{Cucker2001} and less intensive computation of the posterior distribution.
For example, the probability $p(x,t)$ of input $x\in[0,1)$ could be restricted to obey the equation
\begin{align}\label{InitialUserDist}
 \frac{\partial}{\partial t}p(x,t)
     = \frac{\exp\left(-||r_t(x),r(I_t^\tau)||^2 / 2\sigma^2(I_t^\tau) \right) }{\int_0^1 \exp\left(-||r_t(y),r(I_t^\tau)||^2 / 2\sigma^2(I_t^\tau) \right) dy } -1
\end{align}
where $r_t(x)$ is a point in some space $X$,
$||\cdot,\cdot||$ is an appropriately chosen distance function on $X$,
and the learned functions $r(I_t^\tau)$ and $\sigma^2(I_t^\tau)$ correspond roughly to the point indicated by the user and the uncertainty of this indication respectively.
For cases in which $X$ corresponds to the display space and thus $r_t(x)$ is the position at which $x$ is displayed, learning could be accelerated by imposing vertical and/or horizontal reflection symmetries; when $X$ is a circle (as in Fig.\ \ref{layouts}B), rotational symmetries could similarly be applied.
Where such symmetries are only approximate, it may be advisable to gradually reduce the degree to which they are imposed in order to benefit from both faster learning at early times and reduced bias at later times when more data has accumulated.

Often, reasonable initial estimates for the conditional probabilities $p(x,I_t)$ should permit learning to occur simultaneously with task completion and without any prior training.
However, in cases of severe disabilities (\S\ref{sec:disabilities}) or atypical data entry devices (\S\ref{sec:atypical}), learning may have to be initiated with specialized training trials until sufficient accuracy is achieved to make online learning feasible.

\section{Applications}
The generality of RTIAC offers a myriad of potential applications; this section aims to present a few salient and illustrative examples.

\subsection{Simple data entry}
RTIAC can be applied to many tasks in which the user must continuously enter data of a single type.
The obvious example is a RTIAC \textit{Dasher}-type text entry program (Fig.\ \ref{fig:RTIAC_dasher}), which could be opened to enter text files, or could be automatically called whenever the user must enter a character string, be it a URL, file name, email address, etc.
As discussed by the developers of \textit{Dasher} \cite{Ward2000,Ward2002}, such a text entry device, although slower than a traditional keyboard,
is advantageous for use on personal data assistants, by mobility impaired users, or with alphabets not well suited to keyboards.
The expected improved speed of RTIAC relative to IAC data entry may increase the situations in which such a text entry device is useful.
Similar programs could be developed for entering other types of data such as music notation, as discussed in \S\ref{sec:layouts}.

Computer files organized in directories have ordering and tree structure similar to character strings;
therefore, the same type of program could be used for browsing directories and selecting files.
Since the expected time required to select a particular file would depend on the frequency of the file's use rather than its depth within a directory,
such a method of file browsing would eliminate the usual trade-off between quick file access and helpful organization of files into many informative sub-directories.
Analogous remarks are applicable to the selection of commands from a program's menus (\verb|File|, \verb|Edit|, \verb|View|, etc.), 
programs from an application launcher, 
places in a sectioned document such as an HTML documentation file,
or music within an organized collection.

In other cases, for example the selection of icons on a desktop or files in a large directory, there is no natural hierarchical structure among the possible inputs.
The lack of such nested structure removes the advantages of the radial and linear layouts and suggests use of the proportional area layout (Fig. \ref{layouts}C).

\begin{narrowfig}{5in}
\begin{tabbing}
\includegraphics[width=0.5\textwidth,clip=true,trim=0 0 0 0]{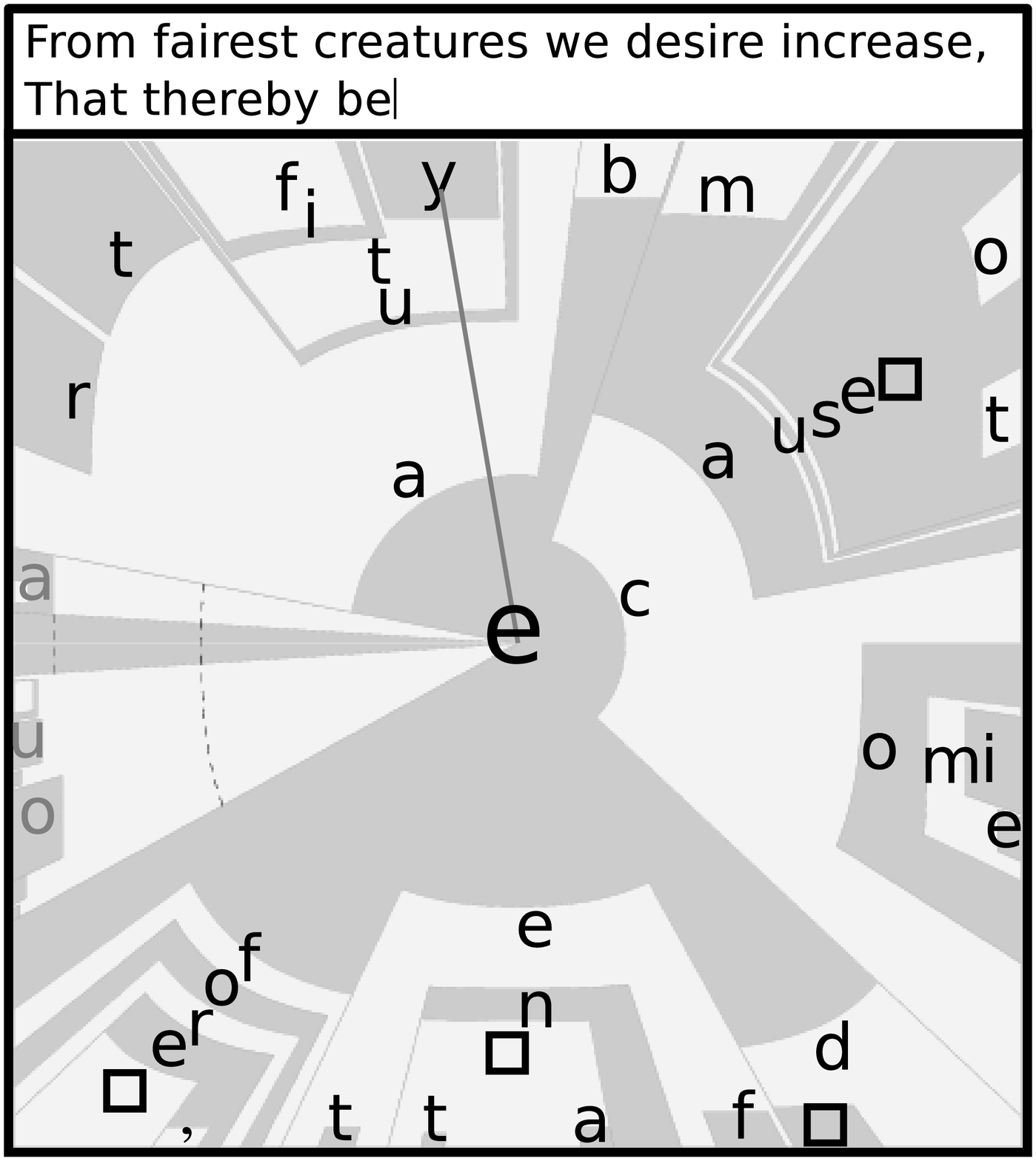}
\end{tabbing}
\caption{\label{fig:RTIAC_dasher} 
A possible text-entry interface based on \textit{Dasher} but implementing RTIAC.
The possible inputs are allotted space such that the probability density of the user choosing a given angle is uniform.
The inputs are ordered as in \textit{Dasher}, beginning above the the gray line segment indicates and wrapping around the circle to end below it.
The character in the middle of the display corresponds to the last entered letter shown above the radial layout.
Characters colored in gray cannot be selected without the deletion of characters already entered;
the number of characters that will be deleted before a gray character is selected is indicated by the number of dashed arcs separating the gray character from the center of the display.
The end of the gray line segment indicates the cursor position.}
\end{narrowfig}

\subsection{Wider scope interaction}\label{sec:wider_scope}
Interfaces using RTIAC can also be applied to tasks of wider scope than the continuous entry of data into a single file or field.
A modest extension of a basic RTIAC text entry device along these lines could some basic navigation and editing commands (\verb|backspace|, \verb|delete|, arrows keys, etc.) among the input options presented to the user.
In addition to allowing the user to quickly manipulate written text as in a normal text editor, the underlying IAC model for such a program could facilitate such predictable input sequences as one backspace followed by a few others to fully eliminate a word.
Commands for movement between cells could also make such an interface suitable for entering data into spreadsheets.

More ambitiously, one could imagine a RTIAC-based user interface for an entire program, multi-program suite, or operating system.
Since users give commands to computers sequentially, an underlying IAC model could be developed in the same way as for strings of sequential alphabet characters.
Such an interface would facilitate transitions between predictably sequential tasks within a single program (e.g.\ printing from a word processor followed by saving the file and exiting the program) or multiple programs (e.g. closing the only open program window followed by either initiating another program, browsing the file directory, or logging out).

\subsection{Suboptimal data entry devices}
RTIAC approaches can be especially helpful when data entry would be otherwise unreliable.
Unlike IAC interfaces, which respond in a fixed way to the user's actions, RTIAC interfaces could learn the specific nature of the entry device's ambiguities and adjust its responses accordingly.
For example, by interpreting the touching of a screen location to increase the probability of nearby inputs rather than to make an immediate selection,
RTIAC interfaces could prevent the frequent errors occurring when users must select between closely spaced icons on a touchscreen or must click on the correct button with an overly sensitive touch-pad mouse.

Hybrid schemes could be developed for interfaces intended for one-time or infrequent users who would not want to learn a new navigation scheme.
For such a text entry device, a standard keyboard layout or array of buttons could be initially displayed with equal-sized buttons;
the user's actions could then serve rescale the two axes as in the proportional area layout, and the original keyboard could be displayed again after each unambiguous key selection.
In this way, the infrequent user need not learn to read nor to navigate the less intuitive layouts of \S\ref{sec:layouts}.

\subsection{Users with disabilities}\label{sec:disabilities}
IAC data entry methods have already benefited disabled computer users by facilitating data entry without the use of a keyboard, for example by tracking gaze direction \cite{Ward2002}.
RTIAC schemes have the potential to provide these benefits in a potentially more effective manner by using the information provided by the user's limited actions to the fullest extent at all times.
With the variety of possible layouts discussed earlier, the advantageous of RTIAC could be enjoyed not only by those capable of performing continuous actions (\S\ref{sec:layouts}), but also by those capable of only timed or untimed discrete actions (\S\ref{sec:single_key}).

Beyond extending the benefits of IAC approaches, RTIAC also provides its own unique benefits to disabled users.
By learning to respond optimally to the actions of individual users, RTIAC interfaces would adapt to users capable of using an input device with only limited efficacy.
Although able to move a mouse or joystick, users with certain movement disorders would be unable to avoid frequent erroneous inputs when using an IAC interface with a preset navigation scheme; 
in contrast, RTIAC interfaces would adapt to these users' poorly controlled motions and then continually update the layout to accurately reflect the information revealed by each action.
Users whose restricted motion prevents them from directing the device in a particular direction would likewise be accommodated by RTIAC interfaces.

\subsection{Atypical input devices}\label{sec:atypical}
Three factors enable RTIAC to better use ambiguous input media, such as live video or sound recording, with which it is less clear which aspects of the user's actions indicate the input intentions.
First, since a user's action serves to update the input probability distribution rather than immediately select an input, the computer is not required to determine the user's intent from a single action, and the user can correct for any misinterpretations with subsequent actions.
Second, the interface need not learn to recognize new actions for each input; rather, it must learn only to associate different actions with different probability measures on $[0,1)$ and thus must only perform a single learning task regardless of the variety of tasks the user wishes to perform.
Third, some of the learning can be transferred to users, who can modify their actions to be better recognized by the interface.

Since RTIAC simply requires learning the associations between the user's actions and the points within the unit interval,
and since it can be used with a wide variety of input devices and combinations thereof,
individual users have extensive freedom in how they interact with an RTIAC interface.
Those wishing to avoid the negative health effects of sedentary computing might train the interface to respond to video-monitored movements providing a vigorous exercise routine.
Ergonomically conscious users may navigate the interface with varied motions to prevent injuries associated with unvarying posture and continuous keyboard or mouse use.
Others may choose to combine normal computer use with practicing a specific activity;
for example, after having trained the interface to associate different pitches with different regions of $[0,1)$, one could practice singing or playing intervals while simultaneously writing a text file.

\section{Discussion}
Drawing from information theory, statistics, and machine learning,
the RTIAC methods presented in this paper are an important step toward intelligent user interfaces that anticipate user's intentions in order to minimize the user's time and effort required for performance of tasks.
By combining the underlying input probability distribution of IAC data entry \cite{Ward2000} with individualized of models how a user's intentions are signaled by his/her actions, RTIAC user interfaces should allow for more rapid data entry,
accommodation of users with disabilities and unreliable input devices,
and greater flexibility in how users interact with computers.
Due to the simplicity of the idea and past research in the areas upon which the RTIAC draws,
there are no substantial obstacles to the creation of such interfaces.

Nevertheless, a number of outstanding issues must be resolved before the advantages of RTIAC can be fully exploited.
An obvious challenge is the development of highly effective interface layouts,
which is especially pertinent in the case of RTIAC interfaces for operating systems and multi-program suites (\S\ref{sec:wider_scope}).
A more subtle question relates to the fact, mentioned in \S\ref{sec:atypical}, that users can modify their behavior in response to mismatches between their intentions and the interpretation of their actions by the RTIAC interface;
how can this co-learning by the user and computer be managed to minimize the time required to complete the learning process and/or to maximize the rate of information transfer in the final state to which the learning converges?
For example, rather than simply learning to respond to the user's actions, the interface could be designed to induce the user to make more varied actions that more effectively distinguish between the possible inputs.
This latter issue is related to the former, since the layout of the user interface will influence the user's actions and thus affect the co-learning procedure.
Identification and development of the most appropriate learning algorithms and hypothesis spaces, which may depend on the interface layout and the medium through which the user communicates, is also important.

\bibliographystyle{acmtrans}
\bibliography{RealTimeInverseArithmeticCoding}
\begin{received}
\end{received}
\end{document}